\documentclass{jfm}

\begin{document}

\newtheorem{lemma}{Lemma}
\newtheorem{corollary}{Corollary}

\newcommand\pb{{^+\!b}}
\newcommand\pc{{^+\!\!c}}
\newcommand\PW{90mm}

\shorttitle{Super compact equation} 
\shortauthor{A. I. Dyachenko et al} 

\title{Super compact equation for water waves}

\author
 {
 A. I. Dyachenko\aff{1,2}
  \corresp{\email{alexd@itp.ac.ru}},
  D. I. Kachulin\aff{2}
  \and 
  V. E.  Zakharov\aff{1,2,3,4}
  }

\affiliation
{
\aff{1}
Landau Institute for Theoretical Physics, 142432, Chernogolovka, Russia
\aff{2}
Novosibirsk State University, 630090, Novosibirsk-90, Russia
\aff{3}
Department of Mathematics, University of Arizona, Tucson, AZ, 857201, USA
\aff{4}
Physical Institute of RAS, Leninskiy prospekt, 53, Moscow, 119991, Russia
}
\maketitle

\begin{abstract}
We derive very simple compact equation for gravity water waves which includes nonlinear wave term 
(\`{a} la NLSE) and advection term (may results in wave breaking). 
\end{abstract}


\section{Introduction}

A potential flow of an ideal incompressible
fluid with free surface in a gravity field is described \citep{Z68} by the following Hamiltonian system:
\begin{equation}\label{I.1}
\frac{\partial \psi}{\partial t} =
-\frac{\delta H}{\delta \eta} \hspace{2cm}
  \frac{\partial \eta}{\partial t} = \frac{\delta H}{\delta \psi}.
\end{equation}
Thereafter we study only the case of one horizontal direction. Now
\begin{eqnarray}
&&\eta = \eta(x,t) \mbox{ - shape of the surface}, \cr
&&\psi = \psi(x,t) = \phi(x,\eta(x,t),t) \mbox{ - potential on the surface}, \cr
&&\phi(x,z,t) \mbox{ - potential inside the fluid}.
\end{eqnarray}
The Hamiltonian $H$ is
\begin{eqnarray}\label{I.2}
H &=& \frac{1}{2}\int\!dx\!\int_{-\infty}^\eta |\nabla\phi|^2dz +  \frac{g}{2}\int \eta^2\!dx
\end{eqnarray}
The potential $\phi(x,z,t)$ satisfies the Laplace equation:
$$
\frac{\partial^2\phi}{\partial^2 x} + \frac{\partial^2\phi}{\partial^2 z} = 0
$$
with the asymptotic boundary conditions:
$$
\frac{\partial\phi}{\partial z}\rightarrow 0,\hspace{20pt}\mbox{at } z \rightarrow -\infty.
$$

\mbox{%
\begin{picture}(260,90)
  \put(0,0){\line(1,0){280}} \put(0,90){\line(1,0){280}}
  \put(0,0){\line(0,1){90}} \put(280,0){\line(0,1){90}}
\put(10,10){\begin{picture}(260,70)%
              \put(0,40){\vector(1,0){240}}
	      \put(130,0){\vector(0,1){70}}
	      \put(230,30){$x$} \put(135,65){$z$}
	      \qbezier[1000](0,0)(20,90)(095,40)
	      \qbezier[1000](095,40)(150,-10)(240,25)
	      \put(7,0){$\eta(x,t)$}
	      \put(190,0){$\psi(x,t)$}
	    \end{picture}}
\end{picture}
}

If the steepness of surface is small, $\eta_x^2 << 1$, the Hamiltonian can be presented by the infinite series
\begin{eqnarray}\label{I.3}
H &=& H_2 + H_3 + H_4 + \dots\cr
H_2 &=& \frac{1}{2}\int (g\eta^2 + \psi \hat k\psi) dx, \cr
H_3 &=& -\frac{1}{2}\int \{(\hat k\psi)^2 -(\psi_x)^2\}\eta dx,\cr
H_4 &=&\frac{1}{2}\int \{\psi_{xx} \eta^2 \hat k\psi + \psi \hat k(\eta \hat k(\eta \hat k\psi))\} dx.
\end{eqnarray}
where $\hat k\psi$ means multiplication by $|k|$ in $k$-space ($|k| = \sqrt{-\frac{\partial^2}{\partial x^2}}$).

Equations (\ref{I.1}) although truncated according to (\ref{I.3}) even for the full 3-D case can be efficiently used
for numerical simulations of water wave dynamics (see, for instance \citep{KPRZ08}). However, they are not convenient 
for analytic study because $\eta(x,t)$ and $\psi(x,t)$ are not "optimal" canonical variables. One can choose better 
Hamiltonian variables by performing a proper canonical transformation. This transformation can be done in two steps.
In the first step we eliminate qubic terms in the Hamiltonian and simplify essentially quartic terms. What we obtain 
as a result of this transformation is so called "Zakharov equation which was widely used in recent years by many 
researchers (see, for instance \citep{CYS80,D94}) of more recent publications \citep{AS11,AS13}. In the second step 
one can "improve" Zakharov equation applying 
appropriate canonical transformation. This "improvement" is possible due to some very special property of the quartic 
Hamiltonian in Zakharov equation. We mean misterious cancellation \citep{DZ94} of nontrivial four-wave interactions.
This cancellation takes place only in one-dimensional case. this cancellation makes possible to replace the "generic" 
Zakharov equation by much more suitable "compact equation", \citep{DZ2011,DZ2012}, which was intensively used as a base 
for both numerical simulations \citep{FD12,FD12a,DKZ13,DKZ14,F14a,F14b,DKZ2015,DKZ2015_1,DKZ2015_2} and analytical proof 
on nonintegrability Of Zakharov equation \citep{DKZ2013a}.

In this paper we discovered that the second step in the canonical transformation is not a unique procedure. One can do it 
by many different ways, obtaining different forms of the compact equation. In this paper we present the most optimal
(by our opinion) version of the compact equation which we call "the super compact equation" for water waves. We present 
also some preliminary results of numerical simulations made by the use of this equation.

\section{Zakharov equation}

Here we briefly recall how to obtain Zakharov equation starting with Hamiltonian (\ref{I.3}). All the detail can be found in 
\cite{Z68, KRS91,ZLF92}.

So, Zakharov equation can be derived in two steps.
\begin{enumerate}
\item[{\bf 1.}] It is convenient to introduce  normal complex variable $a_k$:

\begin{equation}\nonumber
\eta_k =  \sqrt{\frac{\omega_k}{2g}}(a_k+a^*_{-k}) \hspace{.5cm}
\psi_k =  -i\sqrt{\frac{g}{2\omega_k}}(a_k-a^*_{-k})
\end{equation}

\noindent here $\omega_k = \sqrt{gk}$ -is the dispersion law for
the gravity waves, and Fourier transformations $\psi(x)\rightarrow\psi_k$ and $\eta(x)\rightarrow\eta_k$ are defined as follows:
$$
f_k =  \frac{1}{\sqrt{2\pi}}\int f(x)e^{-ikx}dx, \hspace{0.5em}
f(x) = \frac{1}{\sqrt{2\pi}}\int f_k e^{+ikx}dk.
$$
With $a_k$ the Hamiltonian takes the form:
\begin{eqnarray}\nonumber
H_2 & = & \int\!\omega_k a_k a_k^*dk,\cr
H_3 & = & \int\!V^{k}_{k_1 k_2}\{a_k^*a_{k_1}a_{k_2}+a_k a_{k_1}^*a_{k_2}^*\}
\delta_{k-k_1-k_2}\!dkdk_1dk_2\nonumber\cr
&+&\frac{1}{3}\int\!U_{k k_1 k_2}\{a_ka_{k_1}a_{k_2}+a_k^*a_{k_1}^*a_{k_2}^*\}
\delta_{k+k_1+k_2}\!dkdk_1dk_2,\cr
\end{eqnarray}
\begin{eqnarray*}
H_4 = \!\frac{1}{2}\!\int\! W_{k_1k_2}^{k_3k_4}a_{k_1}^*a_{k_2}^*a_{k_3}a_{k_4}\delta_{k_1\!+\!k_2\!-\!k_3\!-\!k_4}\!dk_1\!dk_2\!dk_3\!dk_4 +\cr
+\frac{1}{3}\!\int\! G_{k_1k_2k_3}^{k_4}
(\!a_{k_1}^*\!a_{k_2}^*\!a_{k_3}^*\!a_{k_4}+
c.c.
)\delta_{k_1\!+\!k_2\!+\!k_3\!-\!k_4}\!dk_1\!dk_2\!dk_3\!dk_4 +\cr
+\frac{1}{12}\!\int\! R_{k_1k_2k_3k_4}
(\!a_{k_1}^*\!a_{k_2}^*\!a_{k_3}^*\!a_{k_4}^*\!+\!c.c.)\delta_{k_1\!+\!k_2\!+\!k_3\!+\!k_4}\!dk_1\!dk_2\!dk_3\!dk_4
\end{eqnarray*}
Explicit expressions for coefficients of Hamiltonian are not important here. Nevertheless they can be found  
in \cite{Z98,Z99,DKZ2015}. The motion equations (\ref{I.1}) now take the form: 
\begin{equation}\nonumber
\frac{\partial a_k}{\partial t} + \frac{\delta H}{\delta a^*_k}=0.
\end{equation}
\item[{\bf 2.}] Variables $a_k$ are still not optimal. For transition to better variables $b_k$ one has 
to perform a canonical transformation $a_k\rightarrow b_k$ to cancel all nonresonant cubic and quartic terms in the new 
Hamiltonian. The most economical way to construct the transformation was offered in \citep{ZLF92}. 

By performing the transformation we end up with the Hamiltonian
\begin{eqnarray}\label{HAM_Z}
H=\int\!\omega_k b_k b_k^*\!dk +
\frac{1}{2}\int\! T_{k k_1}^{k_2 k_3} b_k^* b_{k_1}^* b_{k_2}b_{k_3} \delta_{k+k_1-k_2-k_3}dkdk_1dk_2dk_3 + \tilde H
\end{eqnarray}
$\tilde H$ is an infinite series in $b_k, b^*_k$ starting from the fifth order terms. The explicit (and cumbersome) expression for 
$T_{k k_1}^{k_2 k_3}$ can be found in \citep{Z68,Z98,Z99}. The motion equation 
\begin{equation}\label{b_k}
\frac{\partial b_k}{\partial t} + \frac{\delta H}{\delta b^*_k}=0.
\end{equation}
(neglecting $\tilde H$) is the traditional Zakharov equation.
\end{enumerate}

\section{Canonical transformation for Zakharov equation}

A possibility of further simplification of equation (\ref{b_k}) is based on the remarkable fact, established in
\citep{DZ94}. It is the following. Let us consider the resonant condition for four wave interaction
\begin{eqnarray}\label{MAN_RES}
k + k_1 & = & k_2 + k_3,\cr
\omega_k + \omega_{k_1} & = & \omega_{k_2} + \omega_{k_3},
\end{eqnarray}
In 1-D case this system of equations can be resolved as follows:
\begin{eqnarray}\label{Man_2}
k & = & a(1+\zeta)^2, \cr
k_1 & = & a(1+\zeta)^2\zeta^2,\cr
k_2 & = & -a\zeta^2,\cr
k_3 & = & a(1+\zeta+\zeta^2)^2 \hspace{0.5cm} \mbox{here } 0 < \zeta < 1.
\end{eqnarray}
Notice that $kk_1k_2k_3,0$. Now
$$
T_{k k_1}^{k_2 k_3} = F(a,\zeta) = a^3f(\zeta).
$$
Direct calculation shows that 
\begin{equation}\label{ZERO}
f(\zeta) \equiv 0.
\end{equation}
This fact means that "nontrivial" four-wave resonances are absent. However system (\ref{MAN_RES}) has also "trivial" solution:
\begin{equation}\label{Man_1}
k_2=k_1,\hspace{.5cm} k_3=k,\hspace{.5cm}
or\hspace{.5cm}k_2=k,\hspace{.5cm} k_3=k_1,
\end{equation}
We introduce $T_{kk_1}$ (diagonal part) as value of the four-wave coefficient on the trivial manifold (\ref{Man_1}). It was calculated in \citep{Z68} and is equal to:
$$
T_{kk_1} = T_{kk_1}^{kk_1} = \frac{1}{4\pi}|k||k_1|(|k+k_1|-|k-k_1|) =\frac{1}{2\pi}|k||k_1| \mbox{min}(|k|, |k_1|)
$$
Let us introduce $\tilde T_{kk_1}^{k_2k_3}$ as follows:
\begin{eqnarray}\label{T_C}
\!\tilde T_{kk_1}^{k_2k_3} \!=\! \theta(kk_1k_2k_3)\left[
\frac{1}{2}\!(T_{kk_2} \!+\! T_{kk_2} \!+\! T_{k_1k_2} \!+\! T_{k_1k_3})\!-
\!\frac{1}{4}\!(T_{kk}\!+\!T_{k_1k_1}\!+\!T_{k_2k_2}\!+\!T_{k_3k_3})
\right].
\end{eqnarray}
Here $ \theta(k)$ is the step-function. Canonical transformation of the second step has to replace 
cumbersome Zakharov's $T_{kk_1}^{kk_1}$ from (\ref{HAM_Z}) by
much more simple $\tilde T_{kk_1}^{kk_1}$. Obviously their diagonal parts are the same. 

The simple method to construct canonical transformation is based on the fact that a Hamiltonian system keeps at all times 
Hamiltonian properties.It means that transformation $c_k(0)\rightarrow c_k(\tau)$ is canonical.
Let us construct this transformation (as a power series) using some auxiliary Hamiltonian $\tilde H$ (starting from the quartic
term) of the form:
\begin{equation}\nonumber
\tilde H =\frac{1}{2}\int {\bf \tilde B_{kk_1}^{k_2k_3}}c_{k}^*c_{k_1}^*c_{k_2}c_{k_3}\delta_{k+k_1-k_2-k_3}dkdk_1dk_2dk_3+\dots.
\end{equation}
Obviously
\begin{eqnarray}\nonumber
{\bf \tilde B^{k_2k_3}_{kk_1}} = {\bf \tilde B^{k_2k_3}_{k_1k}} =
{\bf \tilde B^{k_3k_2}_{kk_1}} = ({\bf \tilde B^{kk_1}_{k_2k_3}})^*
\end{eqnarray}
Using Taylor series we can express old canonical $b_k(\tau)=c_k(\tau)$ in terms of $c_k(0)$:
\begin{eqnarray*}
c_k(\tau) &=& c_k(0) + \tau \dot c_k(0) +\dots \cr
\dot c_k(0) &=& -i\frac{\delta \tilde H(c_k(0), c^*_k(0))}{\delta c^*_k(0)}
\end{eqnarray*}
and
$$
b_k = c_k -i \int {\bf \tilde B_{kk_1}^{k_2k_3}}c_{k_1}^*c_{k_2}c_{k_3}\delta_{k+k_1-k_2-k_3}dk_1dk_2dk_3+\dots
$$
Now we plug this transformation in the Hamiltonian (\ref{HAM_Z}) of Zakharov equation and get new Hamiltonian:
\begin{eqnarray}\label{HAM_C}
H=\int\!\omega_k c_k c_k^*\!dk &+&\frac{1}{2}\int\! 
\left[ T_{k k_1}^{k_2 k_3} -i(\omega_k+\omega_{k_1}-\omega_{k_2}-\omega_{k_3}){\bf \tilde B^{k_2k_3}_{kk_1}}\right]\times \cr
&\times& c_k^* c_{k_1}^* c_{k_2}c_{k_3} \delta_{k+k_1-k_2-k_3}dkdk_1dk_2dk_3 + \dots
\end{eqnarray}
Coefficient ${\bf \tilde B^{k_2k_3}_{kk_1}}$  of the auxiliary Hamiltonian is also the coefficient of  canonical transformation. It controls the four-wave coefficient $T_{k k_1}^{k_2 k_3}$ in the Hamiltonian of Zakharov equation (\ref{HAM_C}). To replace 
cumbersome $T_{k k_1}^{k_2 k_3}$ by more simple $\tilde T_{k k_1}^{k_2 k_3}$, ${\bf \tilde B^{k_2k_3}_{kk_1}}$ has to be equal to:
\begin{equation}\label{BT1}
{\bf \tilde B_{kk_1}^{k_2k_3}} = i\frac{\tilde T_{k k_1}^{k_2 k_3} - T_{k k_1}^{k_2 k_3}}
{\omega_k+\omega_{k_1}-\omega_{k_2}-\omega_{k_3}}.
\end{equation}
One can check that ${\bf \tilde B_{kk_1}^{k_2k_3}}$ has no singularities at $k+k_1=k_2+k_3$. Indeed in the area where 
$kk_1k_2k_3\le 0$ singularities are canceled in virtue of identity (\ref{ZERO}). In the area where  $kk_1k_2k_3 > 0$ singularities are canceled due to special choice of $\tilde T_{k k_1}^{k_2 k_3}$. Explicit expression for ${\bf \tilde B_{kk_1}^{k_2k_3}}$ was
published in \citep{DLZ95}. By this way we derive the "compact water wave equation".

Due to the absence of nontrivial resonances, waves moving in the same direction do not generate waves moving in the opposite 
direction. Hence we can assume that all $k_{i}>0$. Finally 
\begin{eqnarray}\label{SIMPLE}
\tilde T_{kk_1}^{k_2k_3} &=& \left [ -\frac{1}{8\pi}(kk_2|k-k_2| + kk_3|k-k_3| + k_1k_2|k_1-k_2| + k_1k_3|k_1-k_3|)\right . +\cr
&+& \left .\frac{1}{8\pi}(kk_1(k+k_1) + k_2k_3(k_2+k_3))\right ] \theta(k)\theta(k_1)\theta(k_2)\theta(k_3)
\end{eqnarray}
It corresponds to the following Hamiltonian in $x$-space:
\begin{eqnarray}\label{SPACE_NICE}
H =  \int\!b^*\hat\omega_k bdx +
\frac{1}{2}\int\!\left |\frac{\partial b}{\partial x}\right |^2
\left [\frac{i}{2}\left ( b \frac{\partial b^*}{\partial x} - b^*\frac{\partial b}{\partial x}\right ) -\hat k|b|^2 \right ] dx.
\end{eqnarray}
Here we again went back to using variable $b(x,t)$. The compact equation with the Hamiltonian (\ref{SPACE_NICE}) was used as a base for numerical Simulations in papers \citep{}

\section{Super compact equation}

Now we notice that choice (\ref{T_C}) is not a unique way for introducing a new Hamiltonian. In fact, the conditions imposed on 
$\tilde T_{k k_1}^{k_2 k_3}$ are pretty loose. They area
\begin{enumerate}
\item Symmetry conditions. One must demand that 
$$
{\bf \tilde T^{k_2k_3}_{kk_1}} = {\bf \tilde T^{k_2k_3}_{k_1k}} = {\bf \tilde T^{k_3k_2}_{kk_1}} ={\bf \tilde T^{kk_1}_{k_2k_3}}.
$$
\item The diagonal part must be strictly defined
$$
{\bf \tilde T^{k_2k_3}_{kk_1}} = T_{kk_1} = \frac{1}{4\pi}|k|k_1|(|k+k_1| - |k-k_1|).
$$
\end{enumerate}
Let us choose ${\bf \tilde T^{k_2k_3}_{kk_1}}$ as follows:
\begin{eqnarray}\label{AllPart}
\tilde T_{k_2k_3}^{kk_1} &=& \frac{(kk_1k_2k_3)^{\frac{1}{2}}}{2\pi}\mbox{min}(k,k_1,k_2,k_3)
\times\theta_k\theta_{k_1}\theta_{k_2}\theta_{k_3}\cr
\mbox{min}(k,k_1,k_2,k_3)\!&=&\!\frac{1}{4}(k\!+\!k_1\!+\!k_2\!+\!k_3\!-\!|k-k_2|\!-\!|k-k_3|\!-\!|k_1-k_2|\!-\!|k_1-k_3|)\cr
\mbox{here }\theta_k &-& \mbox{ is the step-function, } \theta_k = \theta(k)
\end{eqnarray}
Now function $b_k$ satisfies the equation:
\begin{equation}\label{AF}
i\dot b_k=\frac{\delta H}{\delta b_k^*} = \omega_k b_k+\frac{k^{\frac{1}{2}}\theta_k}{2\pi}\!\int\!\!\mbox{min}(k,\!k_1,\!k_2\!,k_3\!)
c_{k_1}^*c_{k_2}c_{k_3}\!\delta_{k+k_1-k_2-k_3}dk_1dk_2dk_3
\end{equation}
where 
$$
c_k = k^{\frac{1}{2}}\theta_k b_k
$$
is Fourier-image of analytical (in the upper half-plane) function. Note, nonlinear term in (\ref{AF}) preserve this 
property. Multiplying (\ref{AF}) by $ik^{\frac{1}{2}}$ one can easily get:
\begin{equation}\label{AFF}
\dot c_k + ik\theta_k \left[\frac{\omega_k}{k} c_k+\frac{1}{2\pi}\!\int\!\!\mbox{min}(k,\!k_1,\!k_2\!,k_3\!)
c_{k_1}^*c_{k_2}c_{k_3}\!\delta_{k+k_1-k_2-k_3}dk_1dk_2dk_3\right] = 0
\end{equation}
Expression in square brackets of (\ref{AFF}) is variational derivative of the following Hamiltonian:
\begin{eqnarray}\label{Fourier_C}
H =  \int\!\frac{\omega_k}{k}|c_k|^2dk +\frac{1}{4\pi}\int\!
\mbox{min}(k,\!k_1,\!k_2\!,k_3\!) c_k^*c_{k_1}^*c_{k_2}c_{k_3}\!\delta_{k+k_1-k_2-k_3}dkdk_1dk_2dk_3
\end{eqnarray}
Using following relations between $k$-space and $x$-space
\begin{eqnarray*}\label{K_R}
kc^*_k &\Leftrightarrow& i\frac{\partial}{\partial x}c^*(x),\hspace{1cm}
kc_k \Leftrightarrow -i\frac{\partial}{\partial x}c(x),\\
|k-k_2|c^*_k c_{k_2} &\Leftrightarrow& \hat K (|c(x)|^2),\hspace{1cm} 
(k+k_1)c_k c_{k_1} \Leftrightarrow -i\frac{\partial}{\partial x}(c(x)^2),
\end{eqnarray*}
Relation (\ref{AFF}) is exactly our super compact equation.

Hamiltonian can be written in $x$-space:
\begin{equation}\label{SPACE_C}
H =  \int\!c^*\hat V c \hspace{2pt}dx + \frac{1}{2}\int\!\left [ 
\frac{i}{4}(c^2 \frac{\partial}{\partial x}  {c^*}^2 - {c^*}^2 \frac{\partial}{\partial x} c^2 )-
|c|^2 \hat K(|c|^2)\right ]dx
\end{equation}
Here operator $\hat V$ in K-space is so that $V_k = \frac{\omega_k}{k}$. If along with this to introduce bracket similar to Gardner-Zakharov-Faddeev
\begin{equation}\label{GZF}
\partial^+_x \Leftrightarrow ik\theta_k
\end{equation}
than equation of motion is the following:
\begin{equation}\label{CH}
\frac{\partial c}{\partial t} + \partial^+_x\frac{\delta H}{\delta c^*} = 0.
\end{equation}
Introducing advection velocity
\begin{equation}\label{CV}
{\cal U} = \hat K|c|^2
\end{equation}
taking variational derivative one can write the equation (\ref{CH}) in the form:
\begin{equation}\label{TR}
\frac{\partial c}{\partial t} + i\hat\omega c - 
i\partial^+_x\left (|c|^2\frac{\partial c}{\partial x}\right ) =
\partial^+_x({\cal U}c)
\end{equation}
one can recognize two terms in the equation:
\begin{itemize}
\item nonlinear waving: $i\hat\omega c-i\partial^+_x\left (|c|^2\frac{\partial c}{\partial x}\right )$
\item advection term:  $\partial^+_x({\cal U}c)$.
\end{itemize}

Along with usual quantities such as energy and both momenta equation (\ref{TR})) conserves action or number of waves:
$$
N = \int \frac{|c|^2}{k}dx.
$$

Equation (\ref{TR}) has exact self-similar substitution
$$
c(x,t) = g(t_0-t)^\frac{3}{2}C\left(\frac{x}{g(t_0-t)^2}\right).
$$
Easy to check that $C(\xi)$ satisfies the following equation:
\begin{equation}\label{SS}
\frac{3}{2}C-2\xi\frac{\partial C}{\partial \xi} +i\hat K^{\frac{1}{2}}C -
i\frac{\partial }{\partial \xi}\left(|C|^2\frac{\partial C}{\partial \xi}\right) =
\frac{\partial }{\partial \xi}\left((\hat K|C|^2) C\right)
\end{equation}
where $C(\xi)$ - is dimensionless function which is analytic in the upper half-plane, $\hat K$ -is dimensionless operator.

In $k$-space equation (\ref{AFF}) has the following solution:
\begin{equation}\nonumber
c(k,t) = g^2(t_0-t)^\frac{7}{2}F\left(gk(t_0-t)^2\right)
\end{equation}
Easy to check that dimensionless function $F(\xi)$ satisfies the following equation:
\begin{equation}\label{SSk}
\frac{7}{2}F+2\xi\frac{\partial F}{\partial \xi} = i\xi^{\frac{1}{2}}F +
\frac{i\xi }{2\pi}\int\!\!\mbox{min}(\xi,\!\xi_1,\!\xi_2\!,\xi_3\!)
F^*(\xi_1)F(\xi_2)F(\xi_3)\delta_{\xi+\!\xi_1-\!\xi_2\!-\xi_3\!}
\!d\xi_1\!d\xi_2\!d\xi_3
\end{equation}

\section{Back to $\eta$ and $\psi$}

According to canonical transformation 
$\eta_k$ and $\psi_k$ are power series of $b_k$ 
(or $c_k$) up to the third order:
\begin{equation}\label{EPseries}
\eta_k = \eta_k^{(1)} + \eta_k^{(2)} + \eta_k^{(3)}, \hspace{1cm}
\psi_k = \psi_k^{(1)} + \psi_k^{(2)} + \psi_k^{(3)}.
\end{equation}
Details of the recovering physical quantities  $\eta(x,t)$ and $\psi(x,t)$ are given in \cite{DKZ2015}.
Obviously
\begin{equation}\nonumber
\eta_k^{(1)} = \sqrt{\frac{\omega_k}{2g}}[b_k+b_{-k}^*], \hspace{1cm}
\psi_k^{(1)} = -i\sqrt{\frac{g}{2\omega_k}}[b_k-b_{-k}^*].
\end{equation}
Or
\begin{equation}\nonumber
\eta^{(1)}(x) = \frac{1}{\sqrt{2}g^{\frac{1}{4}}}(\hat k^{\frac{1}{4}}b(x)+\hat k^{\frac{1}{4}}b(x)^*), \hspace{1cm}
\psi^{(1)}(x) = -i\frac{g^{\frac{1}{4}}}{\sqrt{2}}(\hat k^{-\frac{1}{4}}b(x)-\hat k^{-\frac{1}{4}}b(x)^*).
\end{equation}
Operators $\hat k^{\alpha}$ act in Fourier space as multiplication by $|k|^\alpha$.


\begin{eqnarray}\label{EPX}
\eta^{(2)}(x) &=& \frac{\hat k}{4\sqrt{g}}[\hat k^{\frac{1}{4}}b(x) - \hat k^{\frac{1}{4}}b^*(x)]^2,\cr
\psi^{(2)}(x) &=& \frac{i}{2}[\hat k^{\frac{1}{4}}b^*(x)\hat k^{\frac{3}{4}}b^*(x) - \hat k^{\frac{1}{4}}b(x)\hat k^{\frac{3}{4}}b(x)]+ \cr
&+&\frac{1}{2}\hat H[\hat k^{\frac{1}{4}}b(x)\hat k^{\frac{3}{4}}b^*(x) + \hat k^{\frac{1}{4}}b^*(x)\hat k^{\frac{3}{4}}b(x)].
\end{eqnarray}
Here $\hat H$ - is Hilbert transformation with eigenvalue $i\textbf{sign}(k)$.

\section{Numerical Simulation}

\subsection{Breather}

\begin{figure}
\includegraphics[angle=-00,width=\PW]{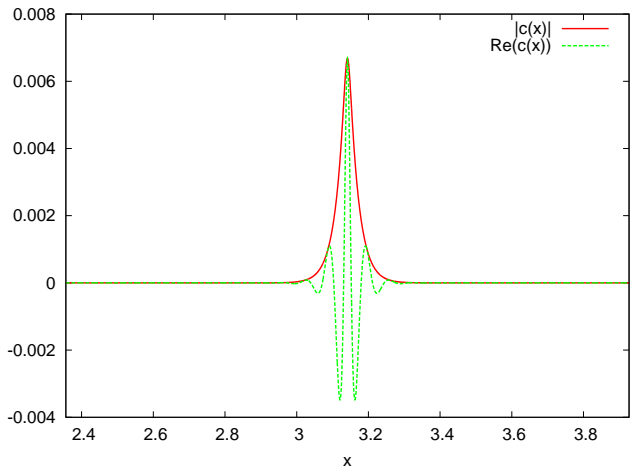}%
\caption{Narrow breather with three crests. $\mbox{Re}(c(x,0))$ and $|c(x,0)|$}
\label{FIG_01}
\end{figure}

\begin{figure}
\includegraphics[angle=-00,width=\PW]{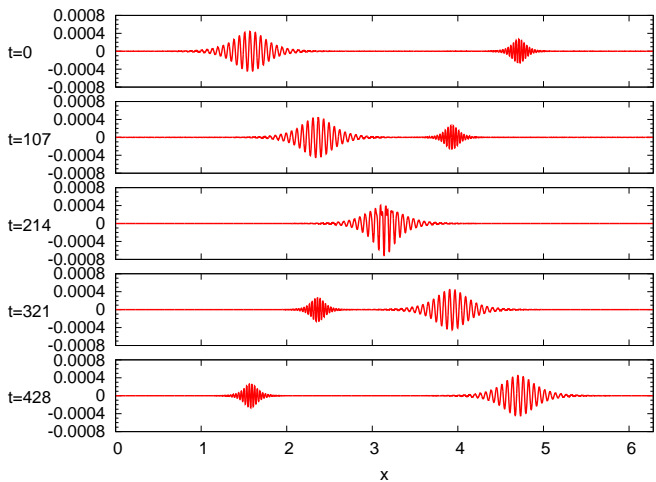}%
\caption{Snapshots of breather collision}
\label{FIG_02}
\end{figure}

Breather is the localized solution of the following type:
\begin{eqnarray*}
c(x,t) = C(x-Vt) e^{i(k_0x - \omega_0 t)} \hspace{0.5cm} or \hspace{0.5cm} c_k = e^{i(\Omega + Vk)t}\phi_k 
\end{eqnarray*}
where $\phi_k$ satisfies the equation:
$$
(\Omega +Vk -\omega_k)\phi_k = \frac{1}{2}\int\! T_{k k_1}^{k_2
k_3}\phi_{k_1}^* \phi_{k_2}\phi_{k_3}
\delta_{k+k_1-k_2-k_3}\!dk_1\!dk_2\!dk_3
$$
It can be found by Petviashvili method
\begin{eqnarray*}
\phi^{n+1}_{k} &=& \frac{NL^{n}_{k}}{M_k}\left [ \frac{<\phi^n \cdot NL(\phi^n)>}{\phi^n \cdot M \phi^n}\right ]^{\gamma}, \hspace{0.5cm} M_k = \Omega +Vk -\omega_k, \cr
NL(\phi^n) &=& -P^{+}\frac{\partial }{\partial x} \left( | \phi^n |^2 \frac{\partial \phi^n}{\partial x} \right)+ iP^{+}\frac{\partial }{\partial x} \left( \hat k \left( |\phi^n|^2\right) \phi^n \right)
\end{eqnarray*}
Breather solution of this equation in the periodic domain $2\pi$ with $k_0=100$ is shown in Fig.\ref{FIG_01}. Breather is very stable structure. Collision of two breathers moving with different velocities (or with $k_0=100$ and $k_0=200$) is shown in 
Fig.\ref{FIG_02}.

\subsection{Modulational instability}

Freak-wave appearing from homogeneous sea with $k_0=100$ and steepness $\mu=0.085$ in the Fig.\ref{FIG_03}:
\begin{figure}
\includegraphics[angle=-00,width=\PW]{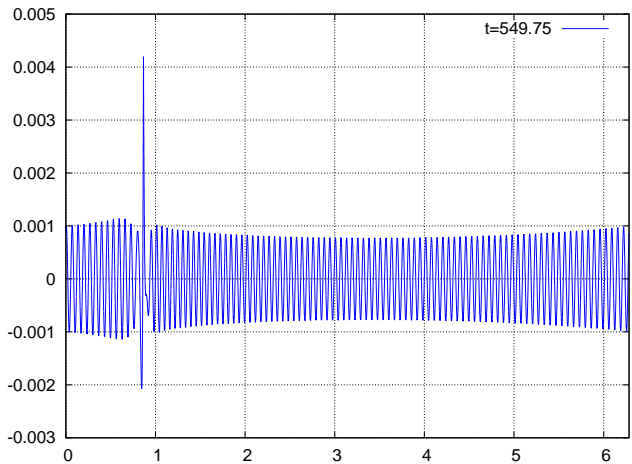}%
\caption{Freak-wave}
\label{FIG_03}
\end{figure}

One can see the beginning of wave breaking in the Fig.\ref{FIG_04}:
\begin{figure}
\includegraphics[angle=-00,width=\PW]{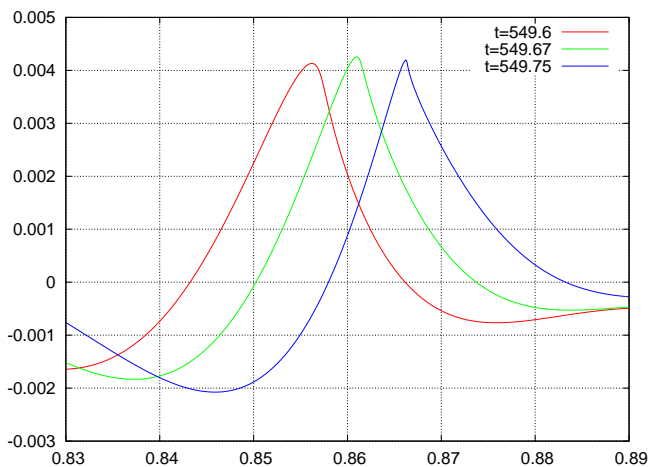}%
\caption{Three snapshots showing beginning of wave breaking}
\label{FIG_04}
\end{figure}

\section{Conclusion}

We derive new compact and elegant form of Hamiltonian and equation for the gravity waves at the surface of deep water.
\begin{itemize}
\item Equation is written for complex normal variable $c(x,t)$ which is analytic function in the upper half-plane
\item Hamiltonian both in $k$-space (\ref{Fourier_C}) and ix $x$-space (\ref{SPACE_C}) is very simple
\item Equation itself is very straightforward consisting of only two terms - nonlinear waves and advection
\item It can be easily implemented for numerical simulation
\end{itemize}

The equation can be generalized for "almost" 2-D waves like KdV is generalized to Kadomtsev-Petviashvili equation:
\begin{eqnarray}\label{SPACE_CXY}
H &=&  \int\!c^*\hat V c \hspace{2pt} dxdy +\cr
&+&\frac{1}{2}\int\!\left [ \frac{i}{4}(c^2 \frac{\partial}{\partial x}  {c^*}^2 -
{c^*}^2 \frac{\partial}{\partial x} c^2 )-|c|^2 \hat K_x(|c|^2)\right ]dxdy
\end{eqnarray}

Here operator $\hat V$ in K-space is $V_{\vec k} = \frac{\omega_{\vec k}}{k_x}$.
\section{Acknowledgments}

This work was supported by Grant 
"Wave turbulence: theory, numerical simulation, experiment" \#14-22-00174 of Russian Science Foundation.

\end{document}